\documentclass[10pt,final,journal]{IEEEtran}

\IEEEoverridecommandlockouts 

\usepackage{def/mypackages}
\usepackage{def/myenvironments}
\usepackage{def/mygraphics}
\usepackage{def/mycolors}
\usepackage{def/myabbr}
\usepackage{def/mymacros}

\begin{document}

\title{Matrix-Valued Measures and Wishart Statistics for Target Tracking Applications
\thanks{This work was performed within the Competence Center SEDDIT, Sensor Informatics and Decision making for the Digital Transformation, supported by Sweden’s Innovation Agency within the research and innovation program Advanced digitalization.

Robin Forsling is with Advanced Programs, Saab AB, Linköping, Sweden, and also with the Department of Electrical Engineering, Linköping University, Linköping, Sweden (e-mail: robin.forsling@liu.se). Simon J. Julier is with the Department of Computer Science, University College London, London, UK. Gustaf Hendeby is with the Department of Electrical Engineering, Linköping University, Linköping, Sweden.
}
}

\author{\IEEEauthorblockN{Robin Forsling, Simon J. Julier, and Gustaf Hendeby} 
}

\maketitle

\IEEEpubid{\begin{minipage}{\textwidth}\ \\[60pt] 
DOI: 10.1109/TAES.2025.3571685 \copyright\xspace 2025 IEEE. Personal use of this material is permitted. Permission from IEEE must be obtained for all other uses, in any current or future media, including reprinting/republishing this material for advertising or promotional purposes, creating new collective works, for resale or redistribution to servers or lists, or reuse of any copyrighted component of this work in other works.
\end{minipage}}


\begin{abstract}
Ensuring sufficiently accurate models is crucial in target tracking systems. If the assumed models deviate too much from the truth, the tracking performance might be severely degraded. While the models are usually defined using multivariate conditions, the measures used to validate them are most often scalar-valued. In this paper, we propose matrix-valued measures for both offline and online assessment of target tracking systems. Recent results from Wishart statistics, and approximations thereof, are adapted and it is shown how these can be incorporated to infer statistical properties for the eigenvalues of the proposed measures. In addition, we relate these results to the statistics of the baseline measures. Finally, the applicability of the proposed measures are demonstrated using two important problems in target tracking: (i) distributed track fusion design; and (ii) filter model mismatch detection.
\end{abstract}



\section{Introduction} \label{sec:introduction}

Target tracking involves estimating the state of a target of interest using noisy sensor measurements. The standard paradigm is model-based target tracking, where sensor models and motion models are combined for tracking the target state over time \cite{Blackman1999MTS}. It is essential for tracking performance that the assumed models are sufficiently correct. If the assumed models deviate too far from the actual underlying models, there is often an unpredictable degradation in the tracking performance.

Developing methodologies and measures for accurate model assessment is still an open challenge \cite{Dunik2020JGCD,Chen2021Fusion,Chen2024TAES}. Model imperfection in target tracking is often evaluated using the \emph{normalized estimation error squared} (\abbrNEES) and the \emph{normalized innovation squared} (\abbrNIS) \cite{Bar-Shalom2001}. Both measures penalize the \emph{mean squared error} (\abbrMSE) weighted by the computed covariance matrix. Hence, they are scale-invariant in contrast to the \abbrMSE. \abbrNEES requires the ground truth to be known and is therefore suitable for offline analyses. Since \abbrNIS can be computed both online and offline it is typically the preferred choice. However, while the models in general are multivariate, both \abbrNEES and \abbrNIS are scalar-valued. Hence, despite their widely spread usage in application areas such as navigation and target tracking, \abbrNEES and \abbrNIS cannot sufficiently address the multivariate relations. Moreover, as pointed out in \cite{Chen2024TAES}, \abbrNEES and \abbrNIS often fail to be useful even for evaluating scalar relations.

In this paper we propose matrix generalizations of the \abbrNEES and \abbrNIS. In particular, by using the eigenvalues of these matrices, different multivariate properties and model imperfections can be examined. We further utilize recent results from Wishart statistics to facilitate the analysis of target tracking systems based on eigenvalue statistics. A few applications\footnote{\matlab code for all developments and applications of this paper is available at: \url{https://github.com/robinforsling/dtt/}.} are used to demonstrate the usage of the proposed matrix-valued measures and the implied statistics.


\section{Related Scalar-Valued Measures} \label{sec:background}

We start with the notation and mathematical preliminaries. Related measures are then reviewed.

\subsection{Notation}

Let $\realsn$ be the set of all $n$-dimensional real-valued vectors. By $\bfA\succeq \bfB$ and $\bfA\succ \bfB$ we denote that the difference $\bfA-\bfB$ is positive semidefinite and positive definite, respectively. The identity matrix of applicable size is given by $\idm$. The expected value and covariance of random vector $\bfz$ are denoted $\EV(\bfz)$ and $\cov(\bfz)$, respectively. 

Let $\bfx_k\in\realsnx$ be an $\nx$-dimensional state at time $k$ to be estimated. An \emph{estimate} of $\bfx_k$ is given by the pair $(\hbfx_k,\bfP_k)$, where $\hbfx_k$ is the state estimate and $\bfP_k\succ\zm$ the covariance computed by the estimator for $\hbfx_k$. Similarly, $(\hbfx_k^i,\bfP_k^i)$ is the estimate computed in the \ith sample or realization, \eg, in a \emph{Monte Carlo} (\abbrMC) simulation. The estimation error is defined as $\tbfx_k = \hbfx_k-\bfx_k$ and $\tbfx_k^i = \hbfx_k^i - \bfx_k$ is the estimation error in the \ith sample. It is assumed that $\hbfx_k$ is unbiased, \ie, $\EV(\tbfx_k)=0$. The matrix $\bfSigma_k=\cov(\tbfx_k)=\EV(\tbfx_k\tbfx_k\trnsp)$ is referred to as the \abbrMSE matrix or the true covariance of the estimation error. Note, we use the same notation for a random variable and a realization of it. 

If $\bfz\sim\calN_m(\bfmu,\bfSigma)$, then $\bfz$ is a Gaussian distributed $m$-dimensional random vector, where $\bfmu=\EV(\bfz)$ and $\bfSigma=\cov(\bfz)$. Moreover, if $\bfz\sim\calN_m(\zm,\idm)$, then $\bfz\trnsp \bfz \sim \chi_m^2$, where $\chi_m^2$ denotes the central chi-squared distribution with $m$ degrees of freedom. Let $\bfZ=\BBM \bfz_1 & \dots & \bfz_n \EBM$ be a $m\times n$ real-valued random matrix, where each column $\bfz_i\sim\calN_m(\zm,\idm)$ is independent and identically distributed (\iid). Then $\bfZ\bfZ\trnsp\sim\calW_m(n,\idm)$ is an $m\times m$ positive semidefinite matrix, where $\calW_m(n,\idm)$ is the real \emph{Wishart distribution} \cite{Wishart1928Biometrika} with $n$ degrees of freedom and covariance parameter $\idm$. The Wishart distribution is the sampling distribution of covariance matrices where the underlying samples are \iid Gaussian random vectors. It is hence relevant when computing sampled covariance matrices from a Gaussian distributed error.


\subsection{Preliminaries}

Two central concepts are \emph{credibility} and \emph{conservativeness}. An estimator of $\bfx_k$ that computes $(\hbfx_k,\bfP_k)$ is \emph{credible} at time $k$ if $\EV(\tbfx_k)=\zm$ and\footnote{It should be pointed out that the probability that a sampled approximation, \eg, an \abbrMC estimator, of $\bfP_k$ equals to $\bfSigma_k$ is zero.}
\begin{equation}
	\bfP_k = \bfSigma_k.
	\label{eq:def:credible-estimator}
\end{equation}
An estimator of $\bfx_k$ that computes $(\hbfx_k,\bfP_k)$ is \emph{conservative} at time $k$ if
\begin{equation}
	\bfP_k \succeq \bfSigma_k.
	\label{eq:def:conservative-estimator}
\end{equation}
In, \eg, \cite{Gao2016TSMC}, the conservativeness criterion is relaxed using the trace operator. An estimator of $\bfx_k$ that computes $(\hbfx_k,\bfP_k)$ is \emph{trace-conservative} at time $k$ if
\begin{equation}
	\trace(\bfP_k) \geq \trace(\bfSigma_k),
	\label{eq:def:trace-conservative-estimator}
\end{equation}
which is a weaker property than conservative \cite{Forsling2022TSP}.

Since $\bfP_k\succ\zm$, it has a unique Cholesky factorization $\bfP_k=\bfL_k\bfL_k\trnsp$, where $\bfL_k$ is lower-triangular and invertible. Moreover, an eigendecomposition of an $n\times n$ symmetric positive semidefinite matrix $\bfS$ is given by
\begin{equation}
	\bfS = \sum_{i=1}^n \lambda_i \bfu_i\bfu_i\trnsp,
	\label{eq:def:eigendecomposition}
\end{equation}
where $\lambda_i=\lambda_i(\bfS)\geq0$ is the \ith eigenvalue of $\bfS$ and $\bfu_i$ is the associated eigenvalue. Note, if unambiguous, for simplicity we use $\lambda_i$ instead of $\lambda_i(\bfS)$ for the \ith eigenvalue of $\bfS$. It is assumed that
\begin{equation}
	\lambdamax = \lambda_1 \geq \dots \geq \lambda_n = \lambdamin.
\end{equation}
The condition in \eqref{eq:def:credible-estimator} is equivalent to
\begin{equation}
	\bfP_k = \bfSigma_k \iff \idm = \bfL_k\inv\bfSigma_k \bfL_k\invtrnsp.
	\label{eq:credible-equivalence}
\end{equation}
Similarly, the condition in \eqref{eq:def:conservative-estimator} is equivalent to
\begin{align}
	\bfP_k \succeq \bfSigma_k
  &\iff \idm \succeq \bfL_k\inv\bfSigma_k \bfL_k\invtrnsp \nonumber \\
  &\iff 1 \geq \lambda_i(\bfL_k\inv\bfSigma_k \bfL_k\invtrnsp), \forall i \nonumber \\
  &\iff 1 \geq \lambdamax(\bfL_k\inv\bfSigma_k \bfL_k\invtrnsp).
	\label{eq:conservativeness-equivalence}
\end{align}


\subsection{Related Work} \label{sec:related-work}

The \abbrNEES is introduced in \cite{Bar-Shalom1983AUT} as a measure for the uncertainty assessment in target tracking algorithms. The \abbrNEES is computed as
\begin{equation}
	\nees_k = \frac{1}{M} \sum_{i=1}^M (\tbfx_k^i)\trnsp (\bfP_k^i)\inv \tbfx_k^i,
	\label{eq:def:nees}
\end{equation}
where $M$ is the number of \abbrMC runs. The \abbrNIS is computed similarly. Let $\tbfy_k$ be the innovation at time $k$, where $\tbfy_k=\bfy_k-\hbfy_k$ is the difference between measurement $\bfy_k$ and predicted measurement $\hbfy_k$ at time $k$, \cf a Kalman filter (\abbrKF, \cite{Kalman1960}). Let $\bfS_k=\bfB_k\bfB_k\trnsp$ be the covariance computed for $\tbfy_k$. Then the \abbrNIS at time $k$ is computed as
\begin{equation}
    \nis_k = \frac{1}{K} \sum_{l=k-K+1}^K \tbfy_l\trnsp \bfS_l\inv\tbfy_l,
	\label{eq:def:nis}
\end{equation}
where $K$ is the number of time steps used.

As pointed out in \cite{Li2001WorkshopETF,Li2002IFAC,Li2006Fusion}, \abbrNEES exhibits a few drawbacks: (i) it penalizes optimism and pessimism asymmetrically; and (ii) it is inconvenient for comparing different estimators' credibility. To overcome these drawbacks the same authors propose the \emph{noncredibility index} (\abbrNCI) defined as
\begin{equation}
	\nci_k
  = \frac{10}{M} \sum_{i=1}^M \log_{10}\left( \frac{(\tbfx_k^i)\trnsp (\bfP_k^i)\inv \tbfx_k^i}{(\tbfx_k^i)\trnsp (\bfSigma_k^i)\inv \tbfx_k^i} \right),
\end{equation}
where $\bfSigma_k^i$ is the \abbrMSE matrix of \ith \abbrMC run. If unknown, $\bfSigma_k^i$ is approximated by
\begin{equation}
	\hat{\bfSigma}_k = \frac{1}{M} \sum_{i=1}^M \tbfx_k^i(\tbfx_k^i)\trnsp.
	\label{eq:Sigmahat}
\end{equation}

In \cite{Bar-Shalom2001} \abbrNEES and \abbrNIS are used to evaluate \emph{filter consistency}. A consistent filter ensures two important properties \cite{Chen2018Fusion}: (i) the error statistics computed by the filter is the same as the true error statistics; and (ii) the filter mixes information obtained from the process with measurement information in an optimal way. To this end, a recent paper \cite{Chen2024TAES} suggests extending \abbrNEES and \abbrNIS by including terms for the second-order moments. This works remarkably well for filter tuning and it has been shown that the extended measures can be integrated into an automatic filter tuning framework \cite{Chen2024TAES}.



The measures mentioned so far---including \abbrNEES, \abbrNIS, and \abbrNCI---are all scalar measures. However, for an estimator to be credible or conservative, certain matrix conditions must be fulfilled. Satisfying the scalar conditions is not sufficient to ensure that the matrix conditions are satisfied. For instance, for a strict evaluation of conservativeness it is necessary to consider semidefinite conditions, \cf \eqref{eq:def:conservative-estimator}. This aspect is briefly addressed in \cite{Li2002IFAC}, where the \emph{credibility interval} is defined as\footnote{For a perfectly credible estimator the credibility interval would reduce to the single value 1.}
\begin{equation}
	[\lambdamin(\bfXi_k),\lambdamax(\bfXi_k)],
	\label{eq:credibility-interval}
\end{equation}
with
\begin{equation}
	\bfXi_k = \bfL_k\inv \bfSigma_k \bfL_k\invtrnsp.
	\label{eq:Gamma}
\end{equation}
To compute $\bfXi_k$, both $\bfSigma_k$ and $\bfP_k=\bfL_k\bfL_k\trnsp$ must be known. A workaround is to approximate these covariances. An unknown covariance $\bfSigma_k=\cov(\tbfx_k)$ can be approximated by $\hat{\bfSigma}_k$ as defined in \eqref{eq:Sigmahat}. Similarly, $\bfP_k$ can be approximated by the mean
\begin{equation}
	\hat{\bfP}_k = \frac{1}{M} \sum_{i=1}^M \bfP_k^i.
	\label{eq:Phat}
\end{equation}
If the system is linear, then $\bfP_k^i=\bfP_k$ for all $i$ and hence $\hat{\bfP}_k=\bfP_k$.

In \cite{Forsling2023Phd}, the \emph{conservativeness index} (\abbrCOIN) is defined by
\begin{equation}
	\coin_k = \lambdamax(\bfL_k\inv\hat{\bfSigma}_k \bfL_k\invtrnsp).
	\label{eq:def:coin}
\end{equation}
The next proposition\footnote{If $\hat{\bfSigma}\neq\bfSigma$, then this is only an approximation. In \cite{Forsling2023Phd} it is assumed that $\bfP_k^i=\bfP_k$. If not, $\bfP_k$ can be approximated by $\hat{\bfP}_k$.} is a direct consequence of \eqref{eq:conservativeness-equivalence}.

\begin{prop}
If $\hat{\bfSigma}_k=\bfSigma_k$, then $(\hbfx_k,\bfP_k)$ is conservative if and only if $\coin_k\leq1$.
\end{prop}

\begin{proof}
An estimate $(\hbfx_k,\bfP_k)$, where $\bfP_k=\bfL_k\bfL_k\trnsp$, is conservative if $\bfP_k\succeq\bfSigma_k$. Hence, if $\hat{\bfSigma}_k=\bfSigma_k$, it follows from \eqref{eq:conservativeness-equivalence} and by definition of $\coin_k$, that
\begin{equation*}
	\bfP_k \succeq \hat{\bfSigma}_k \iff 1 \geq \lambdamax(\bfL_k\inv\hat{\bfSigma}_k \bfL_k\invtrnsp) = \coin_k.
\end{equation*}
\end{proof}

\section{Developing Statistical Measures for Evaluating Matrix-Valued Properties} \label{sec:problem}

We start with a motivating example to illustrate that scalar-valued measures such as \abbrNEES in general fail to evaluate matrix-valued conditions.

\subsection{Motivating Example} \label{sec:motivation}

It is now illustrated how merely looking at \abbrNEES might lead to the conclusion that an estimator is credible or conservative when it in fact is neither. Let
\begin{align*}
    \bfSigma &= \BBM8&1\\1&2\EBM, &
    \bfP &= \BBM8&0\\0&2\EBM.
\end{align*}
Clearly $\bfSigma\neq \bfP$, \ie, the credibility condition is violated. Moreover,
\begin{equation*}
    \bfXi = \BBM1&0.25\\0.25&1\EBM,
\end{equation*}
with eigenvalues $\lambdamin=0.75$ and $\lambdamax=1.25$. Hence, neither the conservativeness condition holds.

\begin{figure}
    \centering
    \begin{tikzpicture}
        \input{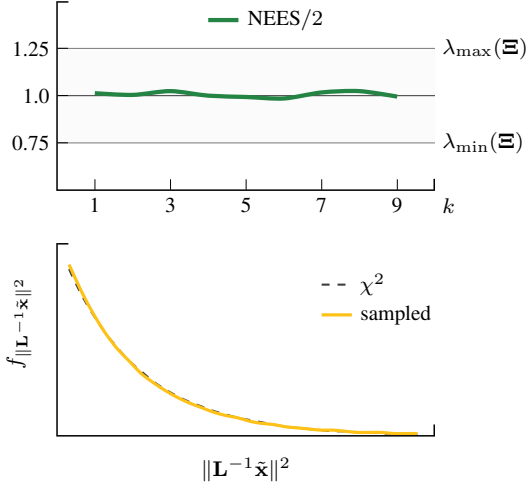}
    \end{tikzpicture}
    \caption{Motivating example. The estimation error is sampled from $\calN_2(\zm,\bfSigma)$ and $\bfP\neq\bfSigma$ is the covariance computed by the estimator. However, from the \abbrNEES it might be concluded that $\bfP=\bfSigma$ despite that $\lambdamin(\bfXi)=0.75$ and $\lambdamax(\bfXi)=1.25$. }
    \label{fig:motivating-example}
\end{figure}

Consider now a stationary setting, where the estimation error $\tbfx_k\sim\calN_2(\zm,\bfSigma)$ and $\bfP=\bfL\bfL\trnsp$ is the covariance computed for $\tbfx_k$ at each $k$. If we sample $\tbfx_k\sim\calN_2(\zm,\bfSigma)$ over independent \abbrMC runs, the \abbrNEES statistics can be computed using \eqref{eq:def:nees}. For a filter consistent estimator we should have $\nees_k=\nx=2$ and that \abbrNEES is $\chi^2$ distributed. The \abbrNEES statistics is plotted in Fig.~\ref{fig:motivating-example} and we see that $\nees_k/2$ is very close to 1. In addition, the sampled probability density function (\abbrPDF) of $\|\bfL\inv\tbfx\|^2$ is computed as a histogram and plotted against the theoretical $\chi^2$ \abbrPDF. By pure inspection, the \abbrNEES statistics is what we would expect for a filter consistent estimator. However, since $\lambdamin(\bfXi)=0.75$ and $\lambdamax(\bfXi)=1.25$ we know that this estimator is not credible nor even conservative.


\subsection{Problem Formulation}

In this paper, the objective is to develop measures that can be used to evaluate matrix-valued conditions, \eg, credibility and conservativeness. This means that we do not only need new measures, but also statistical properties related to these measures.

\section{Proposed Matrix-Valued Statistics} \label{sec:proposed-statistics}

In this section the proposed matrix-valued statistics are presented. We start with a matrix generalization of \abbrNEES which is suitable for offline evaluation of target tracking and data fusion systems. Then a similar generalization of \abbrNIS is proposed that can be used in online applications.

\subsection{The Normalized Estimation Error Squared Matrix} \label{sec:nees-matrix}

The motivating example in the preceding section illustrates that \abbrNEES is not sufficient to evaluate credibility and conservativeness. However, still, the normalized error and innovation are useful tools which we want to generalize to the multivariate case. The \emph{\abbrNEES matrix} is proposed below in Definition~\ref{def:nees-matrix}. It is defined for \abbrMC based simulations and interpreted as the sampled $\bfXi$, \ie, the sampled covariance of the normalized estimation error $\bfL\inv\tbfx$.

\begin{definition}[The \abbrNEES Matrix] \label{def:nees-matrix}
Let $\tbfx_k^i$ be the estimation error of the \ith sample at time $k$. Let $\bfP_k^i=\bfL_k^i(\bfL_k^i)\trnsp\succ\zm$ be the covariance computed by the estimator. The \emph{\abbrNEES matrix} is defined as
\begin{equation}
    \Xihat_k = \frac{1}{M}\sum_{i=1}^M (\bfL_k^i)\inv\tbfx_k^i(\tbfx_k^i)\trnsp (\bfL_k^i)\invtrnsp.
    \label{eq:def:nees-matrix}
\end{equation}
\end{definition}

Note that
\begin{align*}
    \trace(\Xihat_k)
    &= \trace\left( \frac{1}{M} \sum_{i=1}^M (\bfL_k^i)\inv\tbfx_k^i(\tbfx_k^i)\trnsp (\bfL_k^i)\invtrnsp \right) \\
    &= \frac{1}{M} \sum_{i=1}^M \trace\left( (\tbfx_k^i)\trnsp (\bfL_k^i)\invtrnsp(\bfL_k^i)\inv\tbfx_k^i \right) \\
    &= \frac{1}{M} \sum_{i=1}^M (\tbfx_k^i)\trnsp (\bfP_k^i)\inv\tbfx_k^i 
    = \nees_k,
\end{align*}
where it is utilized that $(\bfL_k^i)\invtrnsp(\bfL_k^i)\inv=(\bfP_k^i)\inv$. Let $\lambdamean(\bfA)$ denote the average of the eigenvalues of $\bfA$. If we normalize \abbrNEES using $\nx$
\begin{equation}
    \frac{1}{\nx}\trace(\Xihat_k) = \frac{1}{\nx}\sum_{i=1}^{\nx} \lambda_i(\Xihat_k) = \lambdamean(\Xihat_k),
\end{equation}
which is also referred to as the \emph{average \abbrNEES} \cite{Bar-Shalom2001}.

\subsection{The Normalized Innovation Squared Matrix} \label{sec:nis-matrix}

The \abbrNEES statistic requires knowledge about the true error and is hence used in offline applications where a large number of independent \abbrMC runs are simulated for a particular problem. A statistic that can be computed online, and for single runs, is the \abbrNIS statistics defined in \eqref{eq:def:nis}. In a single run evaluation we average over subsequent time steps instead of \abbrMC runs. The \abbrNIS matrix is proposed in Definition~\ref{def:nis-matrix:online}. It is interpreted as the sampled covariance of the normalized innovation $\bfB\inv\tbfy$.

\begin{definition}[The \abbrNIS Matrix---Single Run Statistics] \label{def:nis-matrix:online}
Let $\tbfy_k\in\realsny$ be the innovation at time $k$. Let $\bfS_k=\bfB_k\bfB_k\trnsp\succ\zm$ be the covariance computed for $\tbfy_k$. The single run \emph{\abbrNIS matrix} is defined as
\begin{equation}
    \Pihat_k = \frac{1}{K} \sum_{l=k-K+1}^k \bfB_l\inv\tbfy_l\tbfy_l\trnsp \bfB_l\invtrnsp.
    \label{eq:def:nis-matrix}
\end{equation}
\end{definition}

Analogously to the \abbrNEES case, we have that
\begin{align*}
    \trace(\Pihat_k)
    &= \trace\left( \frac{1}{K} \sum_{l=k-K+1}^k \bfB_l\inv\tbfy_l\tbfy_l\trnsp \bfB_l\invtrnsp \right) \\
    &= \frac{1}{K} \sum_{l=k-K+1}^k \trace\left( \tbfy_l\trnsp \bfB_l\invtrnsp \bfB_l\inv\tbfy_l \right) \\
    &= \frac{1}{K} \sum_{l=k-K+1}^k \tbfy_l\trnsp \bfS_l\inv \tbfy_l 
    = \nis_k,
\end{align*}
where it is utilized that $\bfS_l\inv=\bfB_l\invtrnsp \bfB_l\inv$. If we divide \abbrNIS by $\ny$
\begin{equation}
    \frac{1}{\ny}\trace(\Pihat_k) = \frac{1}{\ny}\sum_{i=1}^{\ny} \lambda_i(\Pihat_k) = \lambdamean(\Pihat_k).
\end{equation}

In various applications, \eg, filter tuning, it is relevant to compute also the \abbrNIS in an \abbrMC setup. In this case the \abbrNIS matrix is defined as follows.

\begin{definition}[The \abbrNIS Matrix---\abbrMC Statistics] \label{def:nis-matrix:offline}
Let $\tbfy_k^i$ be the innovation of the \ith sample at time $k$. Let $\bfS_k^i=\bfB_k^i(\bfB_k^i)\trnsp\succ\zm$ be the covariance computed for $\tbfy_k^i$. The \abbrMC based \emph{\abbrNIS matrix} is defined as
\begin{equation}
    \Pihatmc_k = \frac{1}{M}\sum_{i=1}^M (\bfB_k^i)\inv\tbfy_k^i(\tbfy_k^i)\trnsp (\bfB_k^i)\invtrnsp.
    \label{eq:def:nis-matrix-mc}
\end{equation}
\end{definition}

\subsection{Test Statistics}

The \abbrNEES matrix and the \abbrNIS matrix are statistics suitable for testing matrix relationships such as credibility, \cf \eqref{eq:def:credible-estimator}, and conservativeness, \cf \eqref{eq:def:conservative-estimator}. In particular, we are interested in the probability distributions of the smallest and largest eigenvalues of $\Xihat$ and $\Pihat$. With such distributions it is, for instance, possible to evaluate if the innovations are samples from a white process with $\bfB_k\inv\tbfy_k\sim\calN(\zm,\idm)$. The \emph{null hypothesis}, in this case, is formulated as
\begin{equation}
    \calH_0 \colon \bfPi_k = \idm,
    \label{eq:def:null-hypothesis}
\end{equation}
which is accepted or rejected using $\lambdamin(\Pihat_k)$ and $\lambdamax(\Pihat_k)$ as test statistics. In the next section we study the marginal distributions of the smallest and largest eigenvalues of Wishart distributed matrices.

\begin{remark}
In this paper we mainly focus on the problem of identifying \emph{if} something is wrong, \eg, model errors, rather than pointing out \emph{how} it is wrong, \eg, which components the model errors affect. This means that we are mainly interested in the eigenvalue statistics. However, in one of the applications we briefly analyze the corresponding eigenvectors which contain information about in which components the models errors contribute.
\end{remark}

\section{Wishart Eigenvalue Statistics} \label{sec:eigenvalue-statistics}

If $\bfL\inv\tbfx\sim\calN_{\nx}(\zm,\idm)$ and $\bfB\inv\tbfy\sim\calN_{\ny}(\zm,\idm)$, then $\Xihat\sim\calW_{\nx}(M,\idm)$ and $\Pihat\sim\calW_{\ny}(K,\idm)$. Hence, we can utilize Wishart statistics to draw conclusions about, credibility, filter consistency, and the models used in a target tracking system.

In this section statistical properties of $\lambdamin(\bfV)$ and $\lambdamax(\bfV)$ are analyzed, where $\bfV\sim\calW_m(n,\idm)$. It is assumed that $n\geq m$ such that $\bfV\succ\zm$.


\subsection{Joint Probability Distribution}

Let $\Gamma(z)$ denote the gamma function, $\gamma(z,a,b)=\int_{a}^{b}t^{z-1}\exp(-t)\,dt$ be the generalized incomplete gamma function, and $r(z,a,b)=\frac{1}{\Gamma(a)}\gamma(z,a,b)$ denote the generalized regularized incomplete gamma function. Define $\Gamma_m(z)=\pi^{m(m-1)/4}\prod_{i=1}^m\Gamma(z-(i-1)/2)$ and $g(z,t)=t^z\exp(-t)$.

Let $\bfV\sim\calW_m(n,\idm)$ and $\lambdamax=\lambda_1\geq\dots\geq\lambda_m=\lambdamin$ be the ordered eigenvalues of $\bfV$. The joint \abbrPDF of $\bflambda = \BBM\lambda_1&\dots&\lambda_m\EBM$ is given by \cite{James1964AMS,Muirhead1982}
\begin{equation}
    f_{\bflambda}(\xi_1,\dots,\xi_m) = K_J \prod_{i=1}^{m} \exp(-\xi_i/2) \xi_i^\alpha\prod_{i<j}^m(\xi_i-\xi_j),
\end{equation}
where $\alpha=(n-m-1)/2$, $\xi_1\geq\dots\geq \xi_m$, and the normalization constant $K_J$ is given by
\begin{equation}
    K_J = \frac{\pi^{m^2/2}}{2^{mn/2}\Gamma_m(m/2)\Gamma_m(n/2)}.
    \label{eq:constant-K1}
\end{equation}

\subsection{Exact Marginal Probability Distributions}

The exact probability that all eigenvalues of $\bfV\sim\calW_m(n,\idm)$ lie within an arbitrary interval is developed in \cite{Chiani2017TINFO}. The cumulative density functions (CDFs) of the smallest and largest eigenvalues of $\bfV$ are then obtained as special cases\footnote{Pioneering work on the marginalization of the extreme eigenvalues of Wishart distributed matrices are found in \cite{James1964AMS,Khatri1964AMS,Edelman1991LAIS,Zanella2009TCOM}.}.

The probability that all eigenvalues of $\bfV\sim\calW_m(n,\idm)$ lie within an interval $[a,b]\subseteq[0,\infty)$ is \cite{Chiani2017TINFO}
\begin{align}
    \psi(a,b)
    &= \Pr\left(a\leq\lambdamin(\bfV) \,,\, \lambdamax(\bfV)\leq b \right) \nonumber \\
    &= K_\lambda \sqrt{\det(\bfA(a,b))},
\end{align}
where
\begin{align}
    K_\lambda
    &= K_J 2^{\alpha m + m(m+1)/2} \prod_{i=1}^m \Gamma(\alpha+i) \nonumber \\
    &= \frac{\pi^{m^2/2}}{\Gamma_m(m/2)\Gamma_m(n/2)} \prod_{i=1}^m \Gamma(\alpha+i),
\end{align}
and $\bfA(a,b)$ is a skew symmetric matrix. A recursive formula for $\psi(a,b)$ is provided in Algorithm~\ref{alg:psi}.

\begin{algorithm}[tb]
	\caption{Probability $\psi(a,b)$ that all eigenvalues of $\bfV\sim\calW_m(n,\idm)$ lie within $[a,b]$ \cite{Chiani2017TINFO}}
	\label{alg:psi}
	\begin{small}
	\begin{algorithmic}[0]
		\Input $m$, $n$, $a$, and $b$
		\State $\bfA = \zm_{m\times m}$ \Comment{$m\times m$ matrix of zeros}
        \State $\alpha_\ell = \alpha+\ell$ \Comment{$\ell$ is an integer}
        \State $K_\lambda=\frac{\pi^{m^2/2}}{\Gamma_m(m/2)\Gamma_m(n/2)} \prod_{i=1}^m \Gamma(\alpha+i)$
        \For{$i=1,\dots,m-1$}
            \For{$j=i,\dots,m-1$}
                \begin{align*}
                    [\bfA]_{i,j+1} &= [\bfA]_{i,j} \\
                    &\quad + \frac{2^{1-\alpha_i-\alpha_j}\Gamma(\alpha_i+\alpha_j)}{\Gamma(\alpha_j+1)\Gamma(\alpha_i)}r(\alpha_i+\alpha_j,a,b) \\
                    &\quad - \frac{g(\alpha_j,a/2)+g(\alpha_j,b/2)}{\Gamma(\alpha_j+1)}r(\alpha_i,a/2,b/2)
                \end{align*}
            \EndFor
        \EndFor
        \If{$m$ is odd}
            \State $\bfc=\zm_{m\times1}$ \Comment{column vector of zeros}
            \For{$i=1,\dots,m$}
                \State $[\bfc]_i = r(\alpha_i,a/2,b/2)$
            \EndFor
            \State $A \leftarrow \BBSM \bfA & \bfc \\ \zm_{1\times m} & 0 \EBSM$
        \EndIf
        \State $\bfA\leftarrow \bfA - \bfA\trnsp$
		\Output $\psi(a,b)=K_\lambda \sqrt{\det(\bfA)}$
	\end{algorithmic}
	\end{small}
\end{algorithm}

With $\psi(a,b)$, the CDFs $F_{\lambdamin}$ and $F_{\lambdamax}$ for the smallest and largest eigenvalues of $\bfV$, respectively, are given by
\begin{subequations} \label{eq:Flambda1m}
    \begin{align}
        F_{\lambdamin}(a) &= \Pr\left(\lambdamin\leq a\right) = 1 - \psi(a,\infty), \\
        F_{\lambdamax}(b) &= \Pr\left(\lambdamax\leq b\right) = \psi(0,b).
    \end{align}
\end{subequations}
since $\psi$ has a positive support.


\subsection{Approximate Marginal Probability Distributions}

The CDFs $F_{\lambdamin}$ and $F_{\lambdamin}$ computed using Algorithm~\ref{alg:psi} are exact. However, for large $m,n$ the asymptotic behavior is often sufficient. In addition, numerical issues might arise when $m$ and $n$ (or $n$ alone) are large\footnote{$m,n$ on the order of $\geq100$ are \emph{large} in this context \cite{Chiani2014JMA}.}. In these situations approximate CDFs are useful. It is known that the smallest and largest eigenvalues converges to a shifted \emph{Tracy-Widom} distribution as $m,n\rightarrow\infty$ \cite{Tracy2009NTMP,Chiani2017TINFO}. Here, we will use the simpler approximations proposed in \cite{Chiani2014JMA,Chiani2017TINFO} which are based upon shifted gamma distributions.

Let $\bfV\sim\calW_m(n,\idm)$ and let $r(z,a)$ be the lower regularized gamma function. Moreover, let $\mu_1$, $\sigma_1^2$, and $s_1$ be the mean, variance, and skewness of the Tracy-Widom distribution\footnote{For details about these parameters, see, \eg, \cite{Tracy2009NTMP}.} of type~1. Define
\begin{align}
    \kappa &= \frac{4}{s_1^2}, &
    \theta &= \frac{\sigma_1s_1}{2}, &
    \rho &= \kappa\theta - \mu_1.
\end{align}
The CDF $F_{\lambdamin}$ is approximated using the result from \cite{Chiani2017TINFO}
\begin{equation}
    \Pr\left(\lambdamin(\bfV) \leq a \right) \approx r\left(\kappa,\frac{\max(0,-a'+\rho)}{\theta}\right),
    \label{eq:approx-cdf-lambdamin}
\end{equation}
where
\begin{subequations}
    \begin{align}
        a' &= \frac{a-\mumin}{\sigmamin}, \\
        \mumin &= \left(\sqrt{n+c_n} - \sqrt{m+c_m} \right)^2, \\
        \sigmamin &= \sqrt{\mumin}\left( \frac{1}{\sqrt{m+c_m}} - \frac{1}{\sqrt{n+c_n}} \right)^{\frac{1}{3}},
    \end{align}
\end{subequations}
and where $c_m$ and $c_n$ are tuning parameters, here set to $c_m=c_n=-1/2$ following \cite{Chiani2014JMA}.

Similarly, $F_{\lambdamax}$ is approximated using \cite{Chiani2014JMA}
\begin{equation}
    \Pr\left(\lambdamax(\bfV) \leq b \right) \approx r\left(\kappa,\frac{\max(0,b'+\rho)}{\theta}\right),
    \label{eq:approx-cdf-lambdamax}
\end{equation}
where
\begin{subequations}
    \begin{align}
        b' &= \frac{b-\mumax}{\sigmamax}, \\
        \mumax &= \left(\sqrt{m+c_m} + \sqrt{n+c_n} \right)^2, \\
        \sigmamax &= \sqrt{\mumax}\left( \frac{1}{\sqrt{m+c_m}} + \frac{1}{\sqrt{n+c_n}} \right)^{\frac{1}{3}}.
    \end{align}
\end{subequations}

\subsection{Expected Values}

Assume that $z$ is a random variable with nonnegative support and CDF $F_z(\zeta)$. Then \cite{Rao1973}
\begin{equation}
    \EV(z) = \int_{0}^{\infty} (1-F_z(\zeta)) \,d\zeta.
\end{equation}
Hence, since both $\lambdamin$ and $\lambdamax$ have positive support only, their expected values are given by
\begin{subequations} \label{eq:ev-lambda}
    \begin{align}
        \EV(\lambdamin) &= \int_{0}^{\infty} (1-F_{\lambdamin}(\xi)) \,d\xi, \\
        \EV(\lambdamax) &= \int_{0}^{\infty} (1-F_{\lambdamax}(\xi)) \,d\xi.
    \end{align}
\end{subequations}

\begin{figure}[tb]
    \centering
    \begin{tikzpicture}[yscale=1.2]
        \input{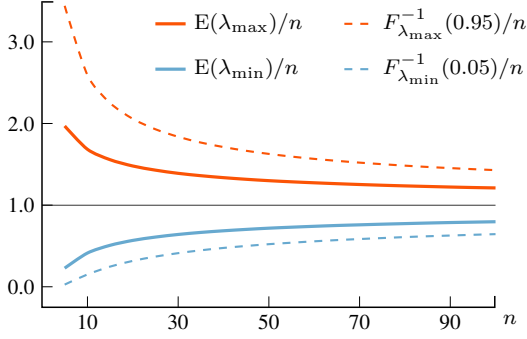}
    \end{tikzpicture}
    \caption{Expected values and inverse CDFs of $\lambdamin$ and $\lambdamax$ as functions of $n$. The curves are normalized with $n$.}
    \label{fig:evlambda-vs-n}
\end{figure}

Let $\bfV\sim\calW_m(n,\idm)$. The expected values of $\lambdamin(\bfV)$ and $\lambdamax(\bfV)$ are plotted in Fig.~\ref{fig:evlambda-vs-n} for $m=3$ and different values of $n$. The inverse CDFs $F_{\lambdamin}\inv$ and $F_{\lambdamax}\inv$ are also plotted, corresponding to one-sided 95\% confidence intervals for $\lambdamin$ and $\lambdamax$, respectively. All curves are normalized with $n$. The curves approaches 1 as $n$ tends to infinity.

\subsection{Relation to $\chi^2$ Statistics}

The Wishart distribution is a multivariate generalization of the $\chi^2$ distribution. Let $\bfZ=\BBM \bfz_1 & \dots & \bfz_n \EBM$, where $\bfz_i\sim\calN_m(\zm,\idm)$ are \iid. Then $\bfZ\bfZ\trnsp\sim\calW_m(n,\idm)$ and
\begin{align*}
    \trace(\bfZ\bfZ\trnsp)
    &= \trace\left( \sum_{i=1}^n \bfz_i\bfz_i\trnsp \right) = \sum_{i=1}^n \trace\left( \bfz_i\bfz_i\trnsp \right) \\
    &= \left( \sum_{i=1}^n \bfz_i\trnsp \bfz_i \right) \sim \chi_{mn}^2.
\end{align*}
Hence, the $\chi^2$ statistics is closely related to the Wishart statistics. However, it is not possible to reconstruct the Wishart statistics from the $\chi^2$ statistics. Still, the $\chi^2$ statistics is relevant when it comes to the evaluation of scalar properties such as trace-conservativeness, \cf \eqref{eq:def:trace-conservative-estimator}, in which case the scalar \abbrNEES can be used.

\subsection{Example: Switching Target Dynamics} \label{sec:example-switching-dynamics}

\begin{figure*}[t]
    \centering
    \begin{tikzpicture}[yscale=1.0]
        \input{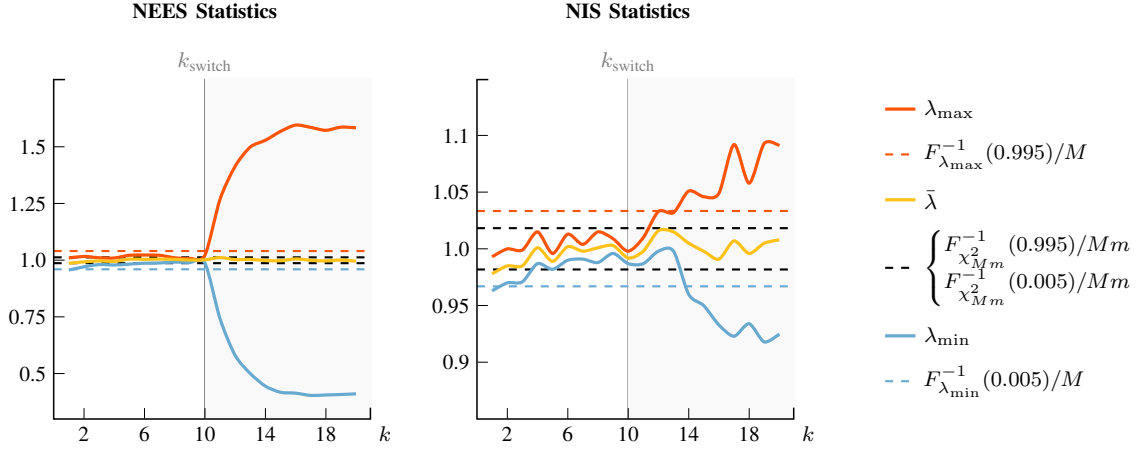}
    \end{tikzpicture}
    \caption{Switching dynamics example. At $\kswitch$, the target dynamics switches from a \abbrCV model with isotropic random accelerations to a \abbrCV model with central random accelerations. The switch is captured by $\lambdamin$ and $\lambdamax$ for both $\Xihat$ and $\Pihatmc$, but not clearly by $\lambdamean$. The confidence intervals given by $F\inv$ have been normalized for comparison reason. }
    \label{fig:change-detection}
\end{figure*}

We will now use a target tracking example to demonstrate the proposed statistics and compare them to their scalar analogs. A target is first tracked using correct models of the dynamics using a \abbrKF. After a certain time, the target dynamics change without changing the models in the \abbrKF. The goal is to be able to detect these changes. To this end, we analyze the \abbrNEES matrix and the \abbrNIS matrix before and after the change in the target dynamics.

Assume two spatial dimensions and let $T_k$ be the sampling time. The target state $x_k$ evolves according to a discrete time (nearly) \emph{constant velocity} (\abbrCV) model
\begin{equation}
    \bfx_{k+1} = \bfF_k\bfx_k + \bfG_k\bfw_k,
    \label{eq:process-model}
\end{equation}
where $\bfw_k\sim\calN_2(\zm,\bfQ_k)$ is the process noise, $\bfQ_k$ the process noise covariance, and\footnote{This corresponds to a sample-and-hold model.}
\begin{align}
    \bfF_k &= \BBM 1&0&T_k&0 \\ 0&1&0&T_k \\ 0&0&1&0 \\ 0&0&0&1 \EBM, &
    \bfG_k &= \BBM \frac{T_k^2}{2}&0 \\ 0&\frac{T_k^2}{2} \\ T_k&0 \\ 0&T_k \EBM.
    \label{eq:ex-switch:Fk-Gk}
\end{align}
A measurement $\bfy_k\in\reals^2$ at time $k$ is given according to the linear measurement model
\begin{equation}
    \bfy_k = \bfH_k\bfx_k + \bfv_k = \BBM 1&0&0&0 \\ 0&1&0&0 \EBM \bfx_k + \bfv_k,
    \label{eq:linear-measurement-model}
\end{equation}
where $\bfv_k\sim\calN_2(\zm,\bfR_k)$ is the measurement noise and $\bfR_k=\sigma_v^2\idm$ the measurement noise covariance.

We will simulate $\bfx_k$ according to \eqref{eq:process-model} using a $\bfQ_k$ that switches at time $\kswitch$. Let $\bfu_k^{\parallel}$ and $\bfu_k^{\perp}$ be two-dimensional unit vectors, where $\bfu_k^{\parallel}$ is longitudinal and $\bfu_k^{\perp}$ is lateral to the target velocity at time $k$. At $k=1,\dots,\kswitch$
\begin{equation}
    \bfQ_k = q^2\BBM1&0\\0&1\EBM,
\end{equation}
and at $k=\kswitch+1,\dots,20$
\begin{equation}
    \bfQ_k = q^2\BBM \bfu_k^{\parallel} & \bfu_k^{\perp} \EBM \BBM 10^{-6}&0\\0&2 \EBM \BBM \bfu_k^{\parallel} & \bfu_k^{\perp} \EBM\trnsp,
\end{equation}
where $q$ is the magnitude of the random acceleration. The target hence first evolves according to isotropic random accelerations and then according to nearly central random accelerations.

The problem is evaluated using \abbrMC simulations. We compute the \abbrNEES matrix $\Xihat$ using Definition~\ref{def:nees-matrix} and the \abbrNIS matrix $\Pihatmc$ using Definition~\ref{def:nis-matrix:offline}. Both are averaged over the \abbrMC simulations for each time $k$. The results are summarized in Fig.~\ref{fig:change-detection}. It is seen that $\lambdamin$ and $\lambdamax$ respond very quickly when the switch occurs. Since the change in dynamics is such that the acceleration decreases in the longitudinal component and increases in the lateral component, $\lambdamin$ falls and $\lambdamax$ rises. This is true for both the \abbrNEES and \abbrNIS statistics. However, $\lambdamean$ is approximately the same after the $\kswitch$. Hence, it would be difficult to observe the change by merely looking at the scalar-valued \abbrNEES and \abbrNIS. Confidence intervals derived from the inverse CDFs are also included. By $F_{\chi_{Mm}^2}\inv$ we denote the inverse CDF of the $\chi^2$ distribution with $Mm$ degrees of freedom.

\section{Target Tracking Applications} \label{sec:applications}

In this section we demonstrate two important applications for the proposed matrix-valued measures. In the first application we consider track fusion design where the task is to choose a track fusion method \emph{offline} evaluation of the \abbrNEES matrix. In the second application we use the \abbrNIS matrix for \emph{online} detection of process model mismatch. In both applications the eigenvalues of \abbrNEES/\abbrNIS matrix are evaluated by utilizing the Wishart statistics presented in the previous section.

\matlab source code for the applications is available at \url{https://github.com/robinforsling/dtt/}. The repository also contains the functionality described in the previous section.

\subsection{Distributed Track Fusion Design}

Track fusion is a type of data fusion. It is an integral part of network-centric target tracking systems\footnote{For instance, target tracking in distributed sensor networks.} where multiple agents track overlapping sets of targets \cite{Forsling2023Phd}. The goal with this example is to illustrate how the \abbrNEES matrix statistics is used to evaluate conservativeness.


\subsubsection{Scenario and Models}

Assume a target tracking scenario where two agents track a common target in two spatial dimensions. The target state $x_k$ is assumed to evolve according to the CV model
\begin{align}
    \bfx_{k+1} &= \bfF_k\bfx_k + \bfw_k, &
    \bfw_k &\sim\calN_4(\zm,\bfQ_k),
\end{align}
where $\bfF_k$ is given in \eqref{eq:ex-switch:Fk-Gk} and
\begin{equation}
    \bfQ_k = q^2 \BBM \frac{T_k^3}{3}&0&\frac{T_k^2}{2}&0 \\ 0&\frac{T_k^3}{3}&0&\frac{T_k^2}{2} \\ \frac{T_k^2}{2}&0&T_k&0 \\ 0&\frac{T_k^2}{2}&0&T_k \EBM,
    \label{eq:track-fusion:Qk}
\end{equation}
with $T_k=1$. At each time $k$ the agents filters their local measurements using a \abbrKF. A nonlinear measurement model is assumed for both agents, where a measurement $\bfy_k^i$ in Agent~$i$ at time $k$ is generated according to
\begin{align}
    \bfy_k^i &= \bfh(\bfx_k,\bfs_k^i) + \bfv_k, &
    \bfv_k &\sim\calN_2(\zm,\bfR_k),
\end{align}
where $\bfs_k^i$ is the position of Agent~$i$, $\bfh(\cdot)$ is a mapping from Cartesian to polar coordinates with origin in $\bfs_k^i$, and $\bfR_k=\diag(\sigma_r^2,\sigma_\phi^2)$ with $\sigma_r^2$ and $\sigma_\phi^2$ denoting the variances of the radial and azimuthal error, respectively. At odd $k$ Agent~1 shares its local track with Agent~2 who fuses the tracks. At even $k$ a local estimates is shared in the opposite direction for track fusion.

Assume now that a local extended Kalman filter (\abbrEKF, \cite{Jazwinski1970}), which only uses local measurements and no track fusion, has already been tuned in a satisfactory way for this particular problem. To utilize a received track a track fusion method is needed. The task is here to select a track fusion method based on performance and uncertainty assessment obtained using an \abbrMC study. For this simple example we consider \emph{covariance intersection} (\abbrCI, \cite{Julier1997ACC}) and the \emph{largest ellipsoid} (\abbrLE, \cite{Benaskeur2002IECON}) method as the two candidate track fusion methods. For implementation details and comparisons, see, \eg, \cite{Forsling2024CSM}. Simulation parameters are summarized in Table~\ref{tab:track-fusion-design:parameters}.

\begin{table}[tb]
	\centering
	\caption{Track fusion design parameters}
	\label{tab:track-fusion-design:parameters}
	\begin{footnotesize}
	\begin{tabular}{cl}
    	\toprule\midrule
    	\textbf{Parameter} & \textbf{Description} \\
    	\midrule
        $M=10\,000$ & number of \abbrMC runs \\
        $q=5$ & process noise parameter [\unitsigmawcvm] \\
        $\sigma_r=100$ & standard deviation of radial uncertainty [\meter] \\
        $\sigma_\phi=2$ & standard deviation of azimuthal uncertainty [$\degrees$] \\
    	\bottomrule
	\end{tabular}
	\end{footnotesize}
\end{table}

\subsubsection{Measures for Estimation Quality}

Let $(\hbfx_k^i,\bfP_k^i)$ denote the local estimate, at time $k$ and in \abbrMC run $i$, after track fusion in one of the agents. Performance is evaluated using the \emph{root mean trace} (\abbrRMT) defined as
\begin{equation}
    \rmt_k = \sqrt{ \frac{1}{M} \sum_{i=1}^M \trace(\bfP_k^i) }.
    \label{eq:def:rmt}
\end{equation}

A robust design of the track fusion must also take into account the uncertainty assessment. To this end we consider conservativeness, which basically means that we want a track fusion method that is able to ensure conservative estimates or at least does not violate the conservativeness property too much. Conservativeness is evaluated using $\lambdamax(\Xihat)$, where $\Xihat$ is the \abbrNEES matrix given in Definition~\ref{def:nees-matrix}. In particular, $\lambdamax(\Xihat)$ is compared to a predetermined value of $F_{\lambdamax}\inv(p)$. For $\bfV\sim\calW_m(M,\idm)$, the confidence parameter $p$ corresponds to the probability that $\lambdamax(V)\leq F_{\lambdamax}\inv(p)$. Hence, since $\Xihat$ is normalized by $M$, we compare $\lambdamax(\Xihat)$ with $F_{\lambdamax}\inv(p)/M$. If $\lambdamax(\Xihat)\leq F_{\lambdamax}\inv(p)/M$, the estimator is considered conservative.  

\subsubsection{Results}

\begin{figure*}[t]
    \centering
    \begin{tikzpicture}[yscale=1]
        \input{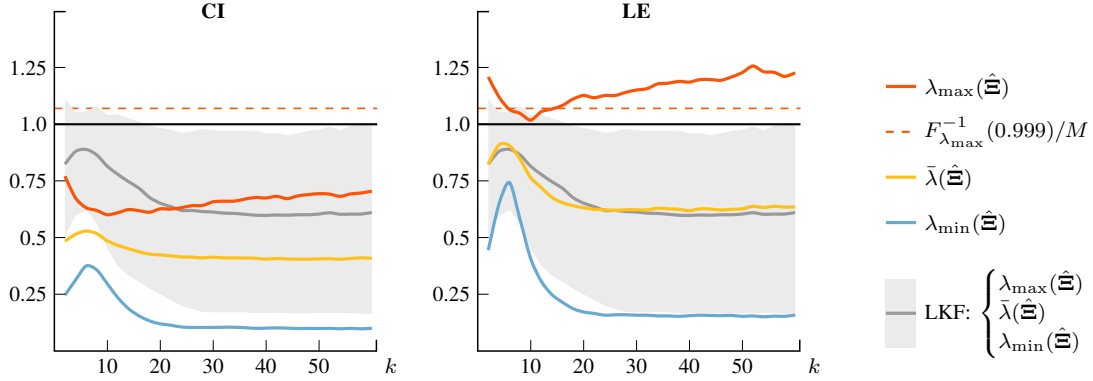}
    \end{tikzpicture}
    \caption{\abbrNEES matrix results for the track fusion design. To evaluate if a track fusion method leads to conservative estimates, $\lambdamax(\Xihat)$ is compared with $F_{\lambdamax}\inv/M$. For convenience, also $\lambdamean(\Xihat)=\nees/\nx$ and $\lambdamin(\Xihat)$ are included. }
    \label{fig:track-fusion-design:nees}
\end{figure*}

\begin{figure}[tb]
    \centering
    \begin{tikzpicture}[yscale=.9]
        \input{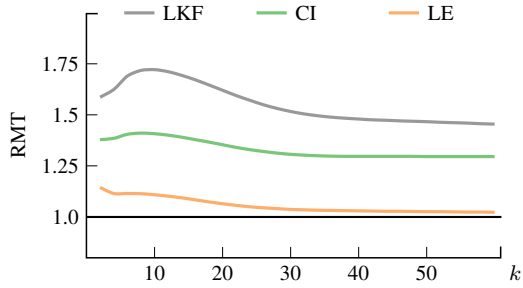}
    \end{tikzpicture}
    \caption{\abbrRMT results for the track fusion design. The \abbrRMT curves have been normalized by the \abbrCRLB.}
    \label{fig:track-fusion-design:rmt}
\end{figure}

The \abbrNEES matrix statistics, computed over all \abbrMC runs, for each $k$ are displayed in Fig.~\ref{fig:track-fusion-design:nees}. The gray area and curve correspond to $\{\lambdamin,\lambdamean,\lambdamax\}$ for the local \abbrEKF (\abbrLKF). Using \abbrCI for track fusion results in conservative but rather pessimistic estimates as $\lambdamax$ is below $1<F_{\lambdamax}\inv(p)/M$. On the other hand, we cannot say that \abbrLE is conservative with respect to (\wrt) the confidence $p$. It is interesting that $\lambdamean$ for \abbrLE does not deviate considerably from the \abbrLKF which by assumption is satisfactorily tuned. Moreover, both are below 1. Hence, if we had only looked at $\lambdamean$, or equivalently $\nees=\nx\lambdamean$, then we would probably have arrived at the conclusion that also \abbrLE is conservative.

The \abbrRMT results are presented in Fig.~\ref{fig:track-fusion-design:rmt}\footnote{Only the results for Agent~1 are presented. The results for Agent~2 are almost identical.}. The curves have been normalized by the \emph{Cram\'{e}r-Rao lower bound} (\abbrCRLB, \cite{Bar-Shalom2001}) such that $\rmt_k=1$ is optimal. It is clearly seen that \abbrLE outperforms \abbrCI \wrt \abbrRMT.

In summary, \abbrLE shows better performance than \abbrCI, but at the cost of not being conservative. The main point is that it requires the \abbrNEES matrix statistics to be able to detect that \abbrLE is not conservative---the \abbrNEES, \cf \eqref{eq:def:nees}, is not sufficient in this case.

\subsection{Filter Model Mismatch Detection}

Using a representative process model is key to the performance of any target tracking system. In practice, the assumed process model used in a tracking filter almost always deviates from the true dynamics of the tracked target. We will now demonstrate how the \abbrNIS matrix can be used online to detect a process model mismatch.

\subsubsection{Scenario and Models}

The considered scenario is similar to the example in Sec.~\ref{sec:example-switching-dynamics}, but without the switching dynamics. In that example, a linear \abbrKF was used to track a single target. The actual target state $x$ evolves in continuous time according to
\begin{equation}
    \dot{\bfx} = \BBM 0&0&1&0 \\ 0&0&0&1 \\ 0&0&0&0 \\ 0&0&0&0 \EBM \bfx + \BBM0\\0\\ \ax \\ \ay \EBM,
    \label{eq:model-mismatch:ct-process}
\end{equation}
where $\dot{\bfx}$ is the time derivative of $\bfx$, $\ax$ and $\ay$ are continuous white noise accelerations along the $x$-axis and $y$-axis, respectively. It is assumed that
\begin{equation}
    \bfQ^a = \cov\left( \BBM \ax \\ \ay \EBM \right) = q^2\BBM \alpha^2 & 0 \\ 0 & 1/\alpha^2 \EBM,
\end{equation}
where $\alpha=2$. Note that, for all $\alpha\neq0$, $\det(\bfQ^a)=q^2$. Discretizing the continuous time model in \eqref{eq:model-mismatch:ct-process} results in $\bfF_k$ according to \eqref{eq:ex-switch:Fk-Gk} and \cite{VanLoan1978TAC}
\begin{equation}
    \bfQ_k = q^2 \BBM \frac{\alpha^2T_k^3}{3}&0&\frac{\alpha^2T_k^2}{2}&0 \\ 0&\frac{T_k^3}{3\alpha^2}&0&\frac{T_k^2}{2\alpha^2} \\ \frac{\alpha^2T_k^2}{2}&0&\alpha^2T_k&0 \\ 0&\frac{T_k^2}{2\alpha^2}&0&\frac{T_k}{\alpha^2} \EBM.
\end{equation}
If $\alpha=1$, then this $\bfQ_k$ reduces to \eqref{eq:track-fusion:Qk}.

The linear sensor model in \eqref{eq:linear-measurement-model} is assumed, where $\bfR_k=\sigma_v^2\idm$. The simulation parameters are summarized in Table~\ref{tab:model-mismatch:parameters}.  

\begin{table}[tb]
	\centering
	\caption{Model mismatch detection parameters }
	\label{tab:model-mismatch:parameters}
	\begin{footnotesize}
	\begin{tabular}{cl}
    	\toprule\midrule
    	\textbf{Parameter} & \textbf{Description} \\
    	\midrule
        $M=10\,000$ & number of \abbrMC runs \\
        $q=10$ & process noise parameter [\unitsigmawcvm] \\
        $\sigma_v=10$ & standard deviation of measurement error [\meter] \\
    	\bottomrule
	\end{tabular}
	\end{footnotesize}
\end{table}

The target is simulated using $\bfQ_k$ with $\alpha=2$ but the \abbrKF uses $\bfQ_k$ with $\alpha=1$. Apart from that, all models used in the \abbrKF are correct. The task is to detect the process model mismatch. Essentially, we want to test the null hypothesis
\begin{equation}
    \calH_0 \colon \bfPi_k = \idm.
\end{equation}
Note, we will not design an actual detection algorithm, but instead compute statistics related to the model mismatch.

\subsubsection{Measures for Model Mismatch Detection}

The \abbrNIS matrix $\Pihat$ is used for online evaluation of the model assumption. We compute an accumulated $\Pihat_k$ in a single run according to
\begin{equation}
    \Pihat_k = \frac{1}{k}\sum_{l=1}^k \bfB_l\inv \tbfy_l\tbfy_l\trnsp \bfB_l\invtrnsp.
\end{equation}

The single runs are evaluated separately using \abbrMC simulations to obtain good statistics. The performance is evaluated using the probability $\poutside$ of detecting a model mismatch. For a certain probability parameter $p\in[0,1]$, we define
\begin{equation}
    \poutside^{\calW} = \Pr\left( (\lambdamin<a_{\lambdamin}) \lor (\lambdamax>b_{\lambdamax}) \right),
    \label{eq:pout-Wishart}
\end{equation}
where $a_{\lambdamin}=F_{\lambdamin}\inv(1-p)$, $b_{\lambdamax}=F_{\lambdamax}\inv(p)$, and $\lor$ denotes logical or. This corresponds to the probability that at least one $\lambda(\Pihat)$ is \emph{outside} a $100(2p-1)\%$ confidence interval. As a reference, we define the corresponding probability for the $\lambdamean(\Pihat)$ statistics accordingly as
\begin{equation}
    \poutside^{\chi^2} = \Pr\left((\lambdamean<a_{\chi^2}) \lor (\lambdamean>b_{\chi^2})\right),
    \label{eq:pout-chi2}
\end{equation}
where $a_{\chi^2}=F_{\chi^2}\inv(1-p)$ and $b_{\chi^2}=F_{\chi^2}\inv(p)$. For a fixed $p$, larger $\poutside$ means a more sensitive detector.

\subsubsection{Complementary Measures}

We also analyze the eigenvectors of $\Pihat$ to investigate if they add any complementary information related to the model mismatch. The idea is that the eigenvectors should contain information about which directions there is a mismatch in the process noise. For instance, in the $x$-axis the actual white noise accelerations have a variance of $q^2\alpha^2$ but the filter is based on $q^2$.

Let $\umaxk$ be the eigenvector associated with $\lambdamax(\Pihat_k)$. By construction, $\Pihat_k$ is computed in a transformed domain due to the $\bfB_k\inv$. We therefore define
\begin{equation}
    \bmaxk = \bfB_k\umaxk.
    \label{eq:bmax}
\end{equation}
Let $\theta_k$ be the angle between $\bmaxk$ and the $x$-axis, which might be both positive and negative, computed in each of the single runs. We compare $\bmaxk$ and the $x$-axis since for the true dynamics the process noise is larger in the $x$-component which should be captured by $\lambdamax$. For each time $k$, we will examine $\bar{\theta}_k$ and $\sigma_{\theta_k}$, corresponding to the mean and standard deviation of $\theta_k$, respectively, obtained by averaging over the \abbrMC runs. Note that, since the eigenvectors are orthogonal, identical results would be obtained by making the corresponding comparison with the eigenvector associated with $\lambdamin(\Pihat_k)$ and the $y$-axis.

\subsubsection{Results}

\begin{figure}[t]
    \centering
    \begin{tikzpicture}[yscale=1.1]
        \input{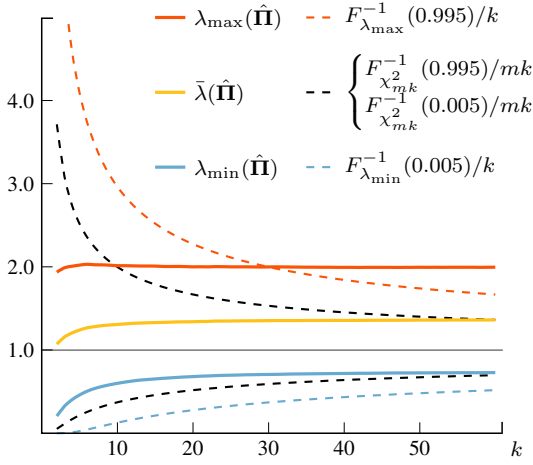}
    \end{tikzpicture}
    \caption{Filter model mismatch detection, where mean values of $\lambdamax$, $\lambdamean$, and $\lambdamin$ are plotted. Dashed lines refer to normalized inverse CDFs of the computed quantities.}
    \label{fig:model-mismatch:lambda}
\end{figure}

Fig.~\ref{fig:model-mismatch:lambda} illustrates the single run statistics with $p=0.995$. The thick solid curves represent mean values of $\lambdamax$, $\lambdamean$, and $\lambdamin$, averaged over the \abbrMC runs for each time $k$. The dashed curves illustrate normalized inverse CDFs under $\calH_0$. We see that $\lambdamax$ crosses $F_{\lambdamax}\inv(0.995)/k$ somewhere around $k=30$. This indicates that there is a significant level of probability to detect the model mismatch. The same cannot be said for $\lambdamean$ which relates to detecting the model mismatch using $\chi^2$ statistics.

The probabilities $\poutside^{\calW}$ and $\poutside^{\chi^2}$ are approximated by their sample means. That is, for each $k$, we average the logical expressions inside $\Pr(\cdot)$ in \eqref{eq:pout-Wishart} and \eqref{eq:pout-chi2} over the \abbrMC runs. The results are plotted in Fig.~\ref{fig:model-mismatch:poutside}, where $\phatoutside^{\calW}$ and $\phatoutside^{\chi^2}$ refer to the sampled approximations of \eqref{eq:pout-Wishart} and \eqref{eq:pout-chi2}, respectively. It is clear that using Wishart statistics the detection performance is significantly improved compared to using the $\chi^2$ statistics.

\begin{figure}[t]
    \centering
    \begin{tikzpicture}[yscale=1.3]
        \input{fig/results_model_mismatch_check_pdet.tex}
    \end{tikzpicture}
    \caption{Filter model mismatch detection, where $\phatoutside^{\calW}$ and $\phatoutside^{\chi^2}$ are sampled approximations of \eqref{eq:pout-Wishart} and \eqref{eq:pout-chi2}, respectively.}
    \label{fig:model-mismatch:poutside}
\end{figure}

\begin{figure}[t]
    \centering
    \begin{tikzpicture}
        \input{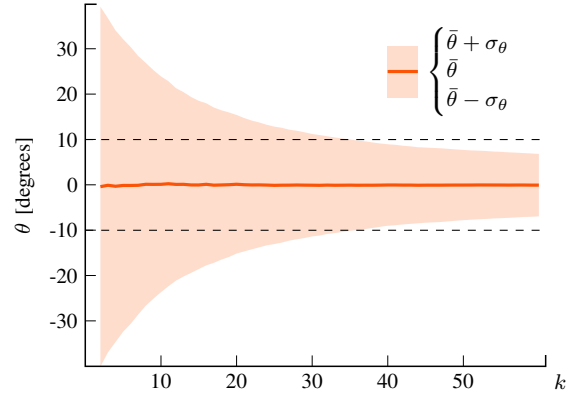}
    \end{tikzpicture}
    \caption{Filter model mismatch detection. The angle $\theta$ represents the deviation of $\bfb_{\max}$ from the $x$-axis.}
    \label{fig:model-mismatch:eigvec}
\end{figure}

The results related to $\bmaxk$ and $\theta_k$ are presented in Fig.~\ref{fig:model-mismatch:eigvec}. These are the single run results which have been averaged over the \abbrMC for easier interpretation. While $\bar{\theta}$ is approximately zero-mean over all $k$, the standard deviation $\sigma_{\theta}$ is initially very high. However, as $k$ increases, $\sigma_{\theta}$ decreases and somewhere between $k=30$ and $k=40$ it becomes less than 10\degree. Hence, it seems like the eigenvectors of $\Pihat$ contain some additional information, although noisy, that can be used to draw conclusions about which components the assumed model fails to match the actual process. This opens up for the possibility to use the eigenvalue statistics to say whether there is a filter model mismatch at all, and then use the eigenvectors to decide how the filter can be retuned, in online applications. However, we consider this to be future work.

\section{Conclusions} \label{sec:conclusions}

We have proposed matrix-valued measures, the \emph{\abbrNEES matrix} and the \emph{\abbrNIS matrix}, with applications to the design and evaluation of target tracking systems. In particular, it has been shown how the eigenvalues of the \abbrNEES and \abbrNIS matrices and the associated eigenvalue statistics can be used to draw conclusions about properties such as credibility, filter consistency, and conservativeness. The applicability of the proposed measures was demonstrated using two target tracking problems: (i) distributed track fusion design; and (ii) filter model mismatch detection.

While the focus of this paper has been on specific target tracking applications, we argue that the proposed measures are useful in essentially all types of estimation problems. For instance, the \abbrNIS matrix can be used to evaluate the correctness of landmark initializations in simultaneous localization and mapping (SLAM). It can also serve as an online computable quality measure for, \eg, localization and decision-making problems in general. It would also be interesting to integrate the proposed measures in an auto-tuning framework such as \cite{Chen2024TAES}. To this end it might be useful to further elaborate on how the eigenvectors of the \abbrNIS matrix can be exploited.

\appendices

%

\bibliographystyle{IEEEtran.bst}
\bibliography{IEEEabrv,bib/myrefs}

\begin{thebibliography}{10}
\providecommand{\url}[1]{#1}
\csname url@samestyle\endcsname
\providecommand{\newblock}{\relax}
\providecommand{\bibinfo}[2]{#2}
\providecommand{\BIBentrySTDinterwordspacing}{\spaceskip=0pt\relax}
\providecommand{\BIBentryALTinterwordstretchfactor}{4}
\providecommand{\BIBentryALTinterwordspacing}{\spaceskip=\fontdimen2\font plus
\BIBentryALTinterwordstretchfactor\fontdimen3\font minus
  \fontdimen4\font\relax}
\providecommand{\BIBforeignlanguage}[2]{{%
\expandafter\ifx\csname l@#1\endcsname\relax
\typeout{** WARNING: IEEEtran.bst: No hyphenation pattern has been}%
\typeout{** loaded for the language `#1'. Using the pattern for}%
\typeout{** the default language instead.}%
\else
\language=\csname l@#1\endcsname
\fi
#2}}
\providecommand{\BIBdecl}{\relax}
\BIBdecl

\bibitem{Blackman1999MTS}
S.~S. Blackman and R.~Popoli, \emph{Design and analysis of modern tracking
  systems}.\hskip 1em plus 0.5em minus 0.4em\relax Norwood, MA, USA: Artech
  House, 1999.

\bibitem{Dunik2020JGCD}
J.~Dun\'{\i}k, O.~Kost, O.~Straka, and E.~Blasch, ``Covariance estimation and
  {G}aussianity assessment for state and measurement noise,'' \emph{J. Guid.,
  Control, Dyn.}, vol.~43, no.~1, pp. 132--139, 2020.

\bibitem{Chen2021Fusion}
Z.~Chen, C.~Heckman, S.~J. Julier, and N.~Ahmed, ``Time dependence in {K}alman
  filter tuning,'' in \emph{Proc. 24th {IEEE} Int. Conf. Inf. Fusion}, Sun
  City, South Africa, Nov. 2021, pp. 1--8.

\bibitem{Chen2024TAES}
Z.~Chen, H.~Biggie, N.~Ahmed, S.~J. Julier, and C.~Heckman, ``Kalman filter
  auto-tuning with consistent and robust {B}ayesian optimization,''
  \emph{{IEEE} Trans. Aerosp. Electron. Syst.}, vol.~60, no.~2, pp. 2236--2250,
  2024.

\bibitem{Bar-Shalom2001}
Y.~Bar-Shalom, X.-R. Li, and T.~Kirubarajan, \emph{Estimation with Applications
  to Tracking and Navigation}.\hskip 1em plus 0.5em minus 0.4em\relax New York,
  NY, USA: John Wiley \& Sons, Ltd, 2001.

\bibitem{Wishart1928Biometrika}
J.~Wishart, ``The generalised product moment distribution in samples from a
  normal multivariate population,'' \emph{Biometrika}, vol. 20A, no. 1/2, pp.
  32--52, 1928.

\bibitem{Gao2016TSMC}
Y.~Gao, X.-R. Li, and E.~Song, ``Robust linear estimation fusion with allowable
  unknown cross-covariance,'' \emph{{IEEE} Trans. Syst., Man, Cybern., Syst.},
  vol.~46, no.~9, pp. 1314--1325, 2016.

\bibitem{Forsling2022TSP}
R.~Forsling, A.~Hansson, F.~Gustafsson, Z.~Sjanic, J.~L\"{o}fberg, and
  G.~Hendeby, ``Conservative linear unbiased estimation under partially known
  covariances,'' \emph{{IEEE} Trans. Signal Process.}, vol.~70, pp. 3123--3135,
  Jun. 2022.

\bibitem{Bar-Shalom1983AUT}
Y.~Bar-Shalom and K.~Birmiwal, ``Consistency and robustness of {PDAF} for
  target tracking in cluttered environments,'' \emph{Automatica}, vol.~19,
  no.~4, pp. 431--437, 1983.

\bibitem{Kalman1960}
R.~E. {Kalman}, ``A new approach to linear filtering and prediction problems,''
  \emph{J. Basic Eng.}, vol.~82, no.~1, pp. 35--45, 1960.

\bibitem{Li2001WorkshopETF}
X.-R. Li, Z.~Zhao, and V.~P. Jilkov, ``Practical measures and test for
  credibility of an estimator,'' in \emph{Proc. Workshop on Estimation,
  Tracking and Fusion: {A} tribute to {Y}aakov {B}ar-{S}halom}, Monterey, CA,
  USA, May 2001.

\bibitem{Li2002IFAC}
------, ``Estimator's credibility and its measures,'' in \emph{Proc. 15th
  Triennial {IFAC} World Congress}, Barcelona, Spain, Jul. 2002.

\bibitem{Li2006Fusion}
X.-R. Li and Z.~Zhao, ``Measuring estimator's credibility: Noncredibility
  index,'' in \emph{Proc. 9th {IEEE} Int. Conf. Inf. Fusion}, Florence, Italy,
  Jul. 2006.

\bibitem{Chen2018Fusion}
Z.~Chen, C.~Heckman, S.~J. Julier, and N.~Ahmed, ``Weak in the {NEES}?:
  {A}uto-tuning {K}alman filters with {B}ayesian optimization,'' in \emph{Proc.
  21st {IEEE} Int. Conf. Inf. Fusion}, Cambridge, UK, Jul. 2018, pp.
  1072--1079.

\bibitem{Forsling2023Phd}
R.~Forsling, ``The dark side of decentralized target tracking: {U}nknown
  correlations and communication constraints,'' {Dissertations. No. 2359},
  Link\"{o}ping University, Link\"{o}ping, Sweden, Nov. 2023.

\bibitem{James1964AMS}
A.~T. James, ``Distributions of matrix variates and latent roots derived from
  normal samples,'' \emph{Ann. Math. Stat.}, vol.~35, no.~2, pp. 475--501, Jun.
  1964.

\bibitem{Muirhead1982}
R.~J. Muirhead, \emph{Aspects of Multivariate Statistical Theory}.\hskip 1em
  plus 0.5em minus 0.4em\relax New York, NY, USA: John Wiley \& Sons, Inc.,
  1982.

\bibitem{Chiani2017TINFO}
M.~Chiani, ``On the probability that all eigenvalues of {G}aussian, {W}ishart,
  and double {W}ishart random matrices lie within an interval,'' \emph{{IEEE}
  Trans. Inf. Theory}, vol.~63, no.~7, pp. 4521--4531, 2017.

\bibitem{Khatri1964AMS}
C.~G. Khatri, ``Distribution of the largest or the smallest characteristic root
  under null hypothesis concerning complex multivariate normal populations,''
  \emph{Ann. Math. Stat.}, vol.~35, no.~4, pp. 1807--1810, 1964.

\bibitem{Edelman1991LAIS}
A.~Edelman, ``The distribution and moments of the smallest eigenvalue of a
  random matrix of {W}ishart type,'' \emph{Linear Algebra Its Appl.}, vol. 159,
  pp. 55--80, 1991.

\bibitem{Zanella2009TCOM}
A.~Zanella, M.~Chiani, and M.~Z. Win, ``On the marginal distribution of the
  eigenvalues of {W}ishart matrices,'' \emph{{IEEE} Trans. Commun.}, vol.~57,
  no.~4, pp. 1050--1060, 2009.

\bibitem{Chiani2014JMA}
M.~Chiani, ``Distribution of the largest eigenvalue for real {W}ishart and
  {G}aussian random matrices and a simple approximation for the
  {T}racy–{W}idom distribution,'' \emph{J. Multivar. Anal.}, vol. 129, pp.
  69--81, 2014.

\bibitem{Tracy2009NTMP}
C.~A. Tracy and H.~Widom, ``The distributions of random matrix theory and their
  applications,'' in \emph{New Trends Math. Phys.}, Dordrecht, Netherlands,
  2009, pp. 753--765.

\bibitem{Rao1973}
C.~R. Rao, \emph{Linear Statistical Inference and its Applications},
  2nd~ed.\hskip 1em plus 0.5em minus 0.4em\relax New York, NY, USA: John Wiley
  \& Sons, Inc., 1973.

\bibitem{Jazwinski1970}
A.~Jazwinski, \emph{Stochastic processes and filtering theory}.\hskip 1em plus
  0.5em minus 0.4em\relax New York, NY, USA: Academic Press, 1970.

\bibitem{Julier1997ACC}
S.~J. Julier and J.~K. Uhlmann, ``A non-divergent estimation algorithm in the
  presence of unknown correlations,'' in \emph{Proc. 1997 Amer. Control Conf.},
  Albuquerque, NM, USA, Jun. 1997, pp. 2369--2373.

\bibitem{Benaskeur2002IECON}
A.~R. {Benaskeur}, ``Consistent fusion of correlated data sources,'' in
  \emph{Proc. 28th Ann. Conf. {IEEE} Ind. Electron. Soc.}, Sevilla, Spain, Nov.
  2002, pp. 2652--2656.

\bibitem{Forsling2024CSM}
R.~Forsling, B.~Noack, and G.~Hendeby, ``A quarter century of covariance
  intersection: {C}orrelations still unknown?'' \emph{{IEEE} Control Syst.
  Mag.}, vol.~44, no.~2, pp. 81--105, 2024.

\bibitem{VanLoan1978TAC}
C.~Van~Loan, ``Computing integrals involving the matrix exponential,''
  \emph{{IEEE} Trans. Autom. Control}, vol.~23, no.~3, pp. 395--404, 1978.

\end{thebibliography}

\end{document}